# Emission of Mitochondrial Biophotons and their Effect on Electrical Activity of Membrane via Microtubules

[1,2,3]Majid Rahnama, [4,5]Jack A. Tuszynski, [6]István Bókkon, [7,8]Michal Cifra,
[1]Peyman Sardar, [1,2,3]Vahid Salari

[1]*Department of Physics, Shahid Bahonar University of Kerman, Kerman, Iran*

[2]*Kerman Neuroscience Research Center (KNRC), Kerman, Iran*

[3]*Afzal Research Institute, Kerman, Iran*

[4]*Department of Experimental Oncology, Cross Cancer Institute, 11560 University Avenue, Edmonton, AB T6G 1Z2, Canada*

[5]*Department of Physics, University of Alberta, Edmonton, T6G 2J1, Canada*

[6]*Doctoral School of Pharmaceutical and Pharmacological Sciences, Semmelweis University, Hungary*

[7]*Institute of Photonics and Electronics, Academy of Sciences of the Czech Republic, Prague, Czech Republic*

[8]*Department of Electromagnetic Field, Faculty of Electrical Engineering, Czech Technical University in Prague, Prague, Czech Republic*

**Abstract**

In this paper we argue that, in addition to electrical and chemical signals propagating in the neurons of the brain, signal propagation takes place in the form of biophoton production. This statement is supported by recent experimental confirmation of photon guiding properties of a single neuron. We have investigated the interaction of mitochondrial biophotons with microtubules from a quantum mechanical point of view. Our theoretical analysis indicates that the interaction of biophotons and microtubules causes transitions/fluctuations of microtubules between coherent and incoherent states. A significant relationship between the fluctuation function of microtubules and alpha-EEG diagrams is elaborated on in this paper. We argue that the role of biophotons in the brain merits special attention.

## 1. Introduction

All living cells of plants, animals and humans continuously emit ultraweak biophotons (ultraweak electromagnetic waves) in the optical range of the spectrum, which is associated with their physiological states and can be measured using special equipment[1]. Neural cells also continuously emit biophotons. The intensity of biophotons is in direct correlation with neural activity, cerebral energy metabolism, EEG activity, cerebral blood flow and oxidative processes [37,50].

---

[1] Updated from: http://www.transpersonal.de/mbischof/englisch/webbookeng.htm, 23 November 2010.

According to Van Wijk et al. [96], there are significant correlations between the fluctuations in biophoton emission and fluctuations in the strength of electrical alpha wave production in the brain. Some unpublished observations suggest that the state of the biophoton field of a person may be connected to the state of the brain as measured by the EEG (e.g., degree of synchronization and coherence) [9]. Certain meditative states characterized by a high degree of coherence in the EEG may well be accompanied by a high degree of coherence in the biophoton field [9], although measurements correlating the coherence of the biophoton field and the EEG readings have not been made yet.

Here, we investigate theoretically the interaction of biomolecules with biophotons taking place within the neurons of the brain. We have adopted a quantum mechanical formalism in an attempt to quantitatively investigate possible connections between the EEG and the biophoton production.

## 2. Biophoton Production Mechanism inside Neurons

During natural metabolic processes taking place in diverse living organism, permanent and spontaneous ultraweak biophoton emission has been observed without any external excitation [1,9,29,36,37,49,50,54,66,70,77,83,89,96,97,107]. The emergence of biophotons is due to the bioluminescent radical and non-radical reactions of Reactive Oxygen Species (ROS) and Reactive Nitrogen Species (RNS), and involves simple cessation of excited states [52,61,102]. The main source of biophotons derives from the oxidative metabolism of mitochondria [94].

Neurons also incessantly emit biophotons [37,50]. Biophoton emission from neural tissue depends on the neuronal membrane depolarization and $Ca^{2+}$ entry into the cells [44]. This biophoton emission can be facilitated by the membrane depolarization of neurons by a high concentration of $K^+$ and can be attenuated by application of tetrodotoxin or removal of extracellular $Ca^{2+}$ [44]. Recently, Sun et al. [82] demonstrated that neurons can conduct photon signals. Moreover, Wang et al. [100] presented the first experimental proof of the existence of spontaneous and visible light induced biophoton emission form freshly isolated rat's whole eye, lens, vitreous humor and retina. They proposed that retinal phosphenes may originate from natural bioluminescent biophotons within the eyes [10,100]. However, the retina is part of the central nervous system. Recently, Bókkon suggested that biophysical pictures may emerge due to redox regulated biophotons in retinotopically organized cytochrome oxidase-rich neural networks during visual perception and imagery within early visual areas [12]. It seems that bioelectronic and biophotonic processes are not independent biological events in the nervous system. Therefore, we conclude that biophoton emission within neurons can be directly correlated with biochemical processes.

However, the term ultraweak bioluminescence (ultraweak biophoton emission) can be misleading, because it may suggest that biophotons are not important for cellular processes. Estimates indicate that for a measured intensity of biophotons, the corresponding intensity of the light field within the organism can be up to two orders of magnitude higher [15,80].

According to Bókkon et al. [11], the real biophoton intensity within cells and neurons can be considerably higher than one would expect from the measurement of ultraweak bioluminescence, which is generally carried out macroscopically several centimeters away from the tissue or cell culture [86]. Moreover, the most significant fraction of natural biophoton intensity cannot be accurately measured because it is absorbed during cellular processes.

Numerous findings have provided evidence of fundamental signal roles of ROS and RNS in cellular processes under physiological conditions [13,14,20,27,33,35,92,95]. There are experimental indications that ROS and RNS are also necessary for synaptic processes and normal brain functions. Free radicals and their derivatives act as signaling molecules in cerebral circulation and are necessary for molecular signaling processes in the brain such as synaptic plasticity, neurotransmitter release, hippocampal long-term potentiation, memory formation, etc. [47,48,85,87,88,99]. Because the generation of ROS and RNS is not a haphazard process, but rather a coordinated mechanism used in signaling pathways, biophoton emission may not be a byproduct of biochemical processes but it can be linked to precise signaling pathways of ROS and RNS. Consequently, regulated generation of ROS and RNS can also produce regulated biophotons within cells and neurons. This means that regulated electrical (redox) signals (*spike-related electrical signals along classical axonal-dendritic pathways*) of neurons can be converted into biophoton signals by various bioluminescent reactions.

Biophotons can be absorbed by natural chromophores such as porphyrin rings, flavinic, pyridinic rings, lipid chromophores, aromatic amino acids, etc. [43,45,57,86]. Mitochondrial electron transport chains contain several chromophores, among which cytochrome oxidase enzymes are most prominent [43,45]. The absorption of biophotons by a photosensitive molecule can produce an electronically excited state. As a result, molecules in electronically excited states often have very different chemical and physical properties compared to their electronic ground states. Regulated biophotons are not dissipated in random manner within cells and neurons, but are absorbed - close to the place where they originated - by chromophores and can excite nearby molecules and trigger/regulate complex signal processes [17]. Thus, absorbed biophotons could have effects on electrical activity of cells and neurons via signal processes.

## 3. Interaction of Mitochondrial Biophotons with Microtubules (MTs)

### *3.1 MTs and Mitochondria*

Microtubular structures have been implicated as playing an important role in the signal and information processing taking place in the human (and possibly animal) brain [32,41,42,55,56]. Vertebrate neurons show typically filamentous mitochondria associated with the microtubules (MTs) of the cytoskeleton, forming together a continuous network (mitochondrial reticulum) [79]. The rapid movements of mitochondria are MT-based and the slower movements are actin-based [91]. MT formation can be regulated by redox-dependent phosphorylations and $Ca^{2+}$ signals. Since the rapid movements of mitochondria are MT-based, mitochondrial trafficking can be organized by redox and $Ca^{2+}$-dependent MT regulation. Moreover, the refractive index of both mitochondria and MTs is higher than the surrounding cytoplasm [86]. Consequently, both the mitochondria and the MTs could act as optical waveguides, i.e. electromagnetic radiation can propagate within their networks [21,42,86]. Associated mitochondrial and MT networks may act as redox and $Ca^{2+}$ regulated organic quantum optical-like fiber systems in neurons. MTs are composed of tubulin dimers.

Tubulin dimer is an intrinsically fluorescent molecule mainly due to 8 tryptophan residues it contains, as can be seen e.g. in the 1TUB structure from the Protein Data Bank [62]. It is well known that the absorption (ca. 280 nm) and fluorescence (ca. 335 nm) wavelength (and intensity) of tryptophan is dependent on the conformation of tubulin. Probing absorption and fluorescence of tubulin is a standard method to determine the polymerization state of MTs. This can be considered one of possible qualitative connections between the fluctuations of MT growth and its corresponding biophoton absorption and emission characteristics. Additionally, there exist other energy states (of both optical and vibrational nature [19,40,65]) which tubulin dimers and the whole MT can support. These states can be excited by energy supply provided by mitochondria [16]. Furthermore, MT polymerization is sensitive to UV [84] and blue light [59] and mitochondria are known to be sources of biophotons corresponding to the same wavelengths [8,34,98], which makes an immediate logical connection. Figure 1 illustrates how mitochondria emit light into MTs.

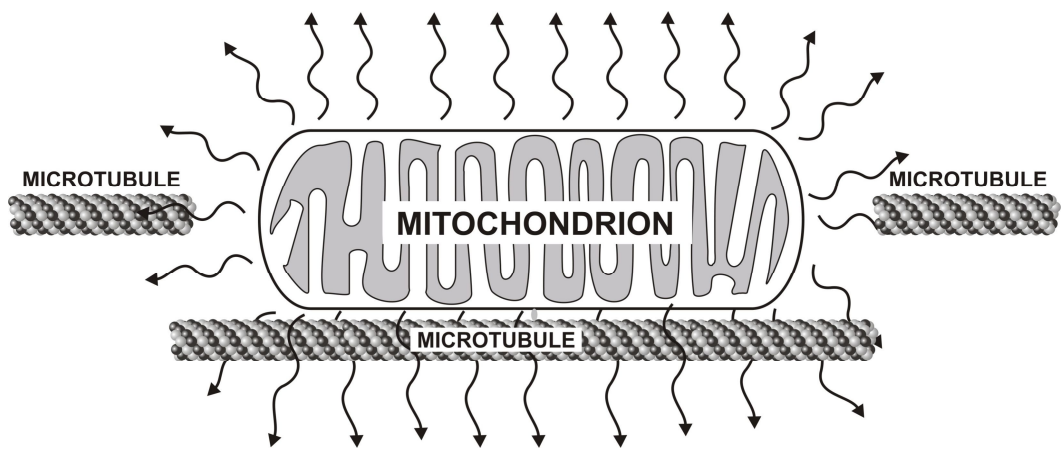

Figure 1 Representation of biophotons produced by mitochondria and the interaction of biophotons with microtubules.

*3.2 Interaction of Biophotons with MTs*

It is worth stressing here that centrioles and cilia, which are complex microtubular structures, are involved in photoreceptor functions in single cell organisms and primitive visual systems. Cilia are also found in all retinal rod and cone cells. The dimensions of centrioles and cilia are comparable to the wavelengths of visible and infrared light [31]. In a series of studies spanning a period of some 25 years G Albrecht-Buehler (AB) demonstrated that living cells possess a spatial orientation mechanism located in the centriole [2,3,4]. This is based on an intricate arrangement of MT filaments in two sets of nine triplets each of which is perpendicular to the other. This arrangement provides the cell with a primitive ''eye'' that allows it to locate the position of other cells within a two to three degree accuracy in the azimuthal plane and with respect to the axis perpendicular to it [2]. He further showed that electromagnetic signals are the triggers for the cells' repositioning. It is still largely a mystery how the reception of electromagnetic radiation is accomplished by the centriole. Another mystery related to these observations is the original electromagnetic radiation emitted by a living cell [94].

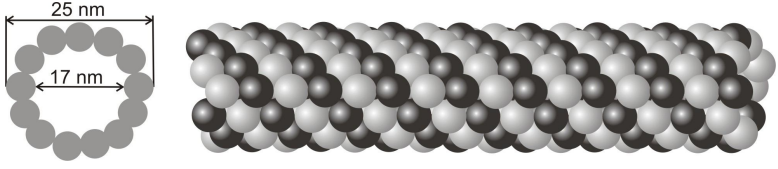

Figure 2 MTs are hollow cylinders composed of protein units called tubulin. The inner diameter of an MT is 17nm and the outer diameter is 25 nm. The lengths of MTs vary widely from nanometers to micrometers. MTs have been considered to act as QED-cavities [55, 56].

Earlier, MTs have been considered as optical cavities [55] with quantum properties [56], capable of supporting only a single mode [41] or perhaps a few widely spaced (in the frequency domain) modes. Our approach is based on a fully quantum mechanical formalism of the Jaynes-Cummings model [39]. MTs are biological hollow cylinders with a 17 nm inner diameter and a 25 nm outer diameter [16], composed of units called tubulin dimers, each of which has the dimensions $4nm \times 8nm \times 6nm$ [55]. Tubulin can be viewed as a typical *two-state* quantum mechanical system, where the dimers couple to conformational changes with $10^{-9} - 10^{-11}$ sec transitions due to electron transitions in hydrophobic pockets, corresponding to an angular frequency in the range $\omega_0 \sim O(10^{10}) - O(10^{12}) Hz$ [55]. Using a first-order-approximation estimate of the quality factor for the MT cavities (i.e. $Q_{MT}$), it has been found that $Q_{MT} \sim O(10^8)$ [55]. High-quality cavities encountered in Rydberg atom experiments dissipate energy on time scales of $O(10^{-3}) - O(10^{-4})$ sec and have Q's which are comparable to $Q_{MT}$ [55]. We consider the tubulin dimer to represent a two-state system with ground $|g\rangle$ and excited $|e\rangle$ states, respectively. Now, we assume that tubulin interacts with a single-mode cavity field of biophotons (the coherent nature of biophotons will be discussed in section 3.4). Here, we introduce the tubulin transition operators $\hat{\sigma}_+, \hat{\sigma}_-$ and $\hat{\sigma}_F$, where $\hat{\sigma}_+ = |e\rangle\langle g|$ is an operator which takes tubulin into the excited state, the operator $\hat{\sigma}_- = |g\rangle\langle e| = \hat{\sigma}_+^\dagger$ takes tubulin into the ground state, and the fluctuation operator $\hat{\sigma}_F = |e\rangle\langle e| - |g\rangle\langle g|$ causes transitions between excited and ground states. We have

$$\begin{aligned}\hat{\sigma}_+|g\rangle = |e\rangle \quad & \hat{\sigma}_+|e\rangle = 0 \\ \hat{\sigma}_-|g\rangle = 0 \quad & \hat{\sigma}_-|e\rangle = |g\rangle\end{aligned} \quad (1)$$

Frequencies of visible light are on the order of $THz$, and as explained before, Wang et al, [100] have detected visible light in the brain as biophotons. Also, transition frequencies in tubulins are on the order of $THz$ [55], so, the interaction between a two-state system (here represented by tubulin) and a single mode quantized field (here represented by biophotons) is given by the total Hamiltonian in the approximation $O(\omega_0) \sim O(\omega)$ [26] as:

$$\hat{H} = \hat{H}_{tubulin} + \hat{H}_{Biophotons} + \hat{H}_{Interation} = \frac{1}{2}\hbar\omega_0\hat{\sigma}_F + \hbar\omega\hat{a}^\dagger\hat{a} + \hbar\lambda(\hat{\sigma}_+\hat{a} + \hat{\sigma}_-\hat{a}^\dagger) \quad (2)$$

where $\hat{H}_{tubulin}$ is the Hamiltonian operator for tubulin, $\hat{H}_{Biophotons}$ is the Hamiltonian operator for biophotons and $\hat{H}_{Interation}$ is the Hamiltonian operator representing the interaction between tubulin and biophotons. Here, $\omega_0$ is the frequency of tubulin transitions and $\omega$ is the frequency of biophotons. $a$ and $a^\dagger$ are annihilation and creation operators, respectively.

$$\hat{a}|n\rangle = \sqrt{n}|n-1\rangle$$
$$\hat{a}^\dagger|n\rangle = \sqrt{n+1}|n+1\rangle \quad (3)$$

where ket $|n\rangle$ is the number state of photons. In Hamiltonian (2) the quantity $\lambda$ is defined as $\lambda = \dfrac{dg}{\hbar}$ where $d$ is the dipole moment of tubulin, $\hbar = \dfrac{h}{2\pi}$ where $h$ is Planck's constant and $g \approx \dfrac{1}{2}\sqrt{\dfrac{\hbar\omega}{\varepsilon V}}$ where $\varepsilon$ is the dielectric constant of the environment in an MT where $\dfrac{\varepsilon}{\varepsilon_0} \sim 80$, in which $\varepsilon_0$ is dielectric constant of vacuum [55], and $V$ is the volume of an MT. Here, the state vector of tubulin is $|tubulin\rangle = c_g(t)|g\rangle + c_e(t)|e\rangle)$ where $c_g(t)$ and $c_e(t)$ are time-dependent coefficients in which $t$ represents time. The state of the field is $|Biophotons\rangle = \sum_{n=0}^{\infty} c_n |n\rangle$ where $c_n$ is a constant coefficient and $|n\rangle$ is the number state of photons. The total state is the tensor product $|\psi(t)\rangle = |tubulin\rangle \otimes |Biophotons\rangle$. Inserting the total state into the Schrödinger's equation $i\hbar \dfrac{d}{dt}|\psi(t)\rangle = \hat{H}_{Interaction}|\psi(t)\rangle$, the result is readily obtained as [26]

$$\rightarrow |\psi(t)\rangle = \sum_{n=0}^{\infty} \left\{ \begin{array}{l} \left[ c_e c_n \cos(\lambda t\sqrt{n+1}) - ic_g c_{n+1}\sin(\lambda t\sqrt{n+1}) \right]|e\rangle|n\rangle \\ + \left[ -ic_e c_{n-1}\sin(\lambda t\sqrt{n}) + ic_g c_n \cos(\lambda t\sqrt{n}) \right]|g\rangle|n\rangle \end{array} \right\} \quad (4)$$

### 3.3 Coherence and Decoherence Problem for the system of MT

The Wu-Austin Hamiltonian has been used to describe the interaction of quantized electromagnetic field with MTs to yield a coherent Froehlich's state of the dipolar biological system. Wu and Austin [104,105,106] proposed a dynamical model containing a biological system composed of electric dipoles with $N$ modes connected to harmonic heat baths representing a quantized electromagnetic source and the surrounding thermal-relaxation bath. The interaction of quantized field with the system of electric dipoles gives a Froehlich coherent state. We believe that the system of neuronal MTs is a good candidate for being properly described by the above Hamiltonian. MTs are composed of tubulin dimers which can be considered as biological electric dipoles.

Previously, one of the concerns of considering coherent states in the brain was due to the fact that the Bose-Einstein condensation happens only at low enough temperatures believed to be lower than body temperature. Recently, Reimers et al. [27] have argued that a very fragile Froehlich coherent state may paradoxically only emerge at very high temperatures and thus there is no possibility for the existence of Froehlich coherent states in biological systems, so quantum models based on the Froehlich coherence should be ruled out. However, it has been shown that there are serious problems in the calculations made by Reimers et al [27] and consequently their conclusions appear to be flawed [75].

Another important problem that remains when considering coherent states for MTs is the so-called *decoherence* problem. The question is "how is it possible for MTs to be in a coherent state while the environment surrounding them is relatively hot, wet and noisy?" Although evidence was found that quantum spin transfer between quantum dots connected by benzene rings (the same structures found in aromatic hydrophobic amino acids) is more efficient at relatively warm temperatures than at absolute zero [63], Tegmark [84] calculated decoherence times for MTs based on the collisions of ions with MTs leading to the decoherence times on the order of:

$$\tau = \frac{D^2 \sqrt{mkT}}{Ngq^2} \approx 10^{-13} s \qquad (5)$$

where $D$ is the tubulin diameter, $m$ the mass of the ion, $k$ Boltzmann's constant, $T$ absolute temperature, $N$ the number of elementary charges in the MT interacting system, $g = \frac{1}{4\pi\varepsilon_0}$ the Coulomb constant and $q$ the charge of an electron. Hagan et al. showed that Tegmark used wrong assumptions for his investigation of MTs. Another main objection to the estimate in (5) is that Tegmark's formulation yields decoherence times that increase with temperature contrary to a well-established physical intuition and the observed behavior of quantum coherent states [74]). In view of these (and other) problems in Tegmark's estimates, Hagan et al. [30] asserted that the values of quantities in Tegmark's relation are not correct and thus the decoherence time should be approximately $10^{10}$ times greater. According to Hagan et al. [30], MTs in neurons of the brain can process information quantum mechanically and they could avoid decoherence via several mechanisms over sufficiently long times for quantum processing to occur. As a result, we conclude that coherent states in MTs are still theoretically possible. Below, we explore the consequences of this conclusion.

### 3.4 Fluctuation Function and Simulation

We begin by representing the state of the MTs as a superposition of the coherent and incoherent (ground) states, and the state of the biophotons as a field composed of n photons. In our approach, transitions involving the coherent-incoherent process determined by a function called the fluctuation function, $F(t)$.

As we explained before, we assume two states for a MT: (a) the ground state (incoherent without energy pumping), and (b) an excited state (coherent with energy pumping). We reformulate our previous calculations for the system of MTs and investigate its interaction with biophotons. First, prior to the interaction, MTs may be in one of the two states: the ground state or the coherent state.

The state of MTs can be written in the form of a superposition of the ground state and coherent state. MTs vibrate in different frequency modes before the energy pumping occurs. This state can be written in the form $|n_1\rangle|n_2\rangle|n_3\rangle.... \equiv |n_1,n_2,n_3,...\rangle = |\{n_j\}\rangle$ in which each $|n_i\rangle$ is a special frequency mode state of a tubulin in an MT. When a quantized electromagnetic field is pumped into the MT, according to Froehlich's theory [22,23,24], this leads to the occupation of one frequency mode with a higher energy. This higher energy state is a coherent state. For simplicity in our calculations, we assume the ground state of the MT to be equivalent to the ket $|0\rangle$. Now, the state of MT can be written as the superposition of the ground state and coherent state,

$$|\psi(0)\rangle_{MT} = c_g |0\rangle + c_e |z\rangle \qquad (6)$$

where $|z\rangle$ is our representation for the coherent state. It is written as $|z\rangle = e^{-\frac{|z|^2}{2}} \sum_{n=0}^{\infty} \frac{z^n}{\sqrt{n!}} |n\rangle$, where the coherent state $|z\rangle$ satisfies in $\hat{a}|z\rangle = z|z\rangle$ where $\hat{a}$ is the annihilation operator. We now investigate its interaction with N biophotons. The state of biophotons is considered as

$$|\phi(0)\rangle_{biophotons} = \sum_{n=0}^{\infty} c_n |n\rangle \qquad (7)$$

and the total state is

$$|\psi(0)\rangle = |\psi(0)\rangle_{MT} \otimes |\varphi(0)\rangle_{biophotons} \qquad (8)$$

After solving the Schrödinger's equation we have the time-dependent wave function as

$$|\psi(t)\rangle = \sum_{n=0}^{\infty} \left\{ \begin{array}{l} \{[c_e c_n \cos(\lambda t \sqrt{n+1}) - i c_g c_{n+1} \sin(\lambda t \sqrt{n+1})]|z\rangle \\ + [-i c_e c_{n-1} \sin(\lambda t \sqrt{n}) + i c_g c_n \cos(\lambda t \sqrt{n})]|0\rangle\} \end{array} \right\} |n\rangle$$

In this case, the MT system first occupies the ground state. It means $c_g = 1$, and $c_e = 0$. After the substitution of these coefficients, the state becomes

$$|\psi(t)\rangle = \sum_{n=0}^{\infty} \left\{ \begin{array}{l} \{[-i c_{n+1} \sin(\lambda t \sqrt{n+1})]|z\rangle \\ + [i c_n \cos(\lambda t \sqrt{n})]|0\rangle\} \end{array} \right\} |n\rangle \qquad (9)$$

Written in another form the state is

$$|\psi(t)\rangle = |\psi_g(t)\rangle |0\rangle + |\psi_e(t)\rangle |z\rangle \qquad (10)$$

where

$$|\psi_e(t)\rangle = -i \sum_{n=0}^{\infty} c_n \sin(\lambda t \sqrt{n}) |n-1\rangle \qquad (11)$$

$$|\psi_g(t)\rangle = i\sum_{n=0}^{\infty} c_n \cos(\lambda t\sqrt{n})|n\rangle \qquad (12)$$

Now, we introduce the fluctuation operator $\sigma_F$ as

$$\sigma_F = |e\rangle\langle e| - |g\rangle\langle g| = |z\rangle\langle z| - |0\rangle\langle 0| \qquad (13)$$

To determine when the MT is in the coherent state and when in the ground state, we use the fluctuation function. We define the fluctuation function as

$$F(t) = \langle\psi|\sigma_F|\psi\rangle$$

This function determines the rate of transitions between the coherent state and the ground state. Eventually, the fluctuation function for this state is

$$F(T,z) = \sum_{n=0}^{\infty} c_n^2 \cos^2(\sqrt{n}T)\left[e^{-|z|^2} - 1\right] + \sum_{n=0}^{\infty} c_n^2 \sin^2(\sqrt{n}T)\left[e^{-\frac{|z|^2}{2}} + e^{-|z|^2}\right] \qquad (14)$$

where $T = \lambda t$. We refer to the parameter $T$ as "scaled time". Here, $c_n$ is the coefficient describing the biophoton field and thus $|c_n|^2$ is the probability of detection of biophotons. We may infer information about coherent light from photo-count statistics (PCS) [26]. One calculates the probability $p(n, \Delta t)$ of registering n photons in preset time interval $\Delta t$ by recording the number of photons during the measurement. A fully coherent field satisfies a Poissonian distribution $p(n) = \frac{N^n e^{-N}}{n!}$ for all times $\Delta t \succ 0$, where $N$ is the average number of detected photons per time interval $\Delta t$. The fact that a coherent field satisfies a Poissonian distribution is rooted in quantum theory. The amplitude and photon numbers are not simultaneously measurable with arbitrary certainty. However, definite and robust experimental proof is often complicated due to very small photon counts in $\Delta t$. Biophotons have been considered coherent by other researchers and there are some claims of experimental observations of coherent biophotons [6,7,9, 66, 67], but in general the claim of biophoton coherence still requires a concrete proof. However, here we hypothetically assume the coherence property of biophotons and investigate its consequences. Thus for biophotons we have

$$c_n = e^{-\frac{|\alpha|^2}{2}} \frac{\alpha^n}{\sqrt{n!}}$$

where the normalization implies that $|\alpha|^2 = N$ [26], keeping in mind that $N$ is the average number of biophotons. Thus, $c_n^2 = e^{-N}\frac{N^n}{n!}$. Finally, the fluctuation function becomes

$$F(T,z) = \sum_{n=0}^{\infty} e^{-N}\frac{N^n}{n!}\left[1 - \cos^2(\sqrt{n}T) + \sin^2(e^{-|z|^2/2}\sqrt{n}T)\right]$$

The 3D diagrams of the fluctuation function are plotted in terms of the scaled time and coordinate z, for different numbers of biophotons N. According to earlier calculations [11], at least 100 biophotons can be produced within each human visual neuron per second during visual perception. So, a few biophotons per second are probably simultaneously present within the 1-10 micrometer length of an MT. Here, we have considered different values for the numbers of biophotons around an MT such as N=2, 5, 7 and 10 and plotted 3D and 2D diagrams of the fluctuation function in Figures 3 and 4, respectively.

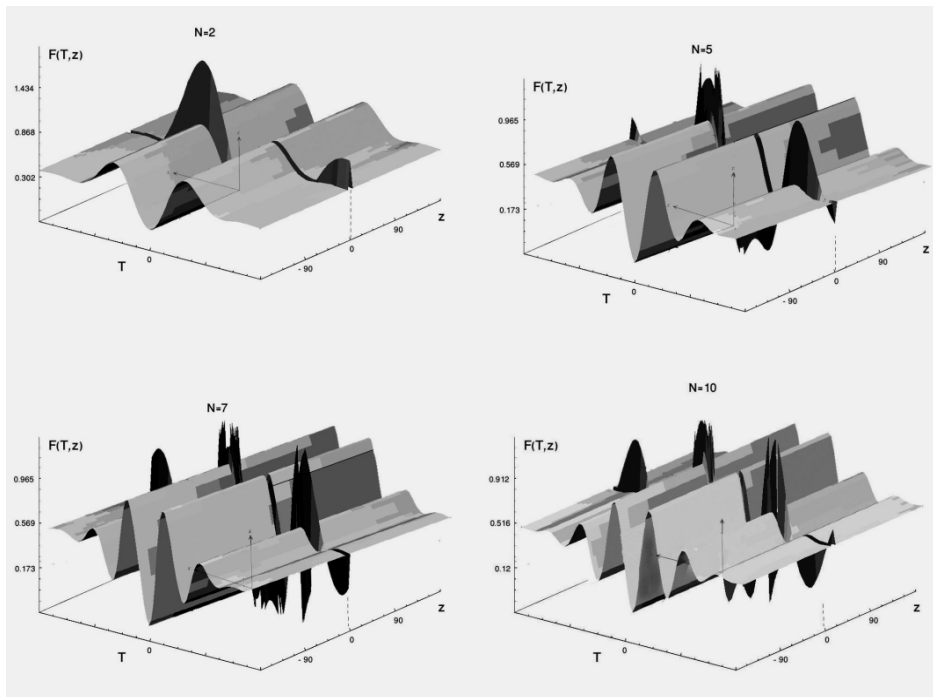

Fig. 3. 3D diagrams of the fluctuation function for different numbers of biophotons N. It is seen that the maximum of fluctuation is around z=1. MT is initially in the ground state

As introduced above, F(T,z) is the fluctuation function which determines the transition rate of biomolecules between the coherent states and the ground state. When in coherent states, the biophotons are absorbed via biomolecules and when in the ground state, the biophotons are absorbed via the vacuum. In this model, the information can be restored from the vacuum, and conscious states can be repeated as before. Then, the information can be sent back to the memory site again via emission of biophotons. This cycle can be repeated an arbitrary number of times. We have plotted the F(T,z) function for different values of its variables. According to 3D diagrams in Figure 3, we let the values near 1 for z, since the main fluctuations are in this area. In the next diagrams the scaled times are evaluated as *T=$10^4$* and *T=$10^5$*.

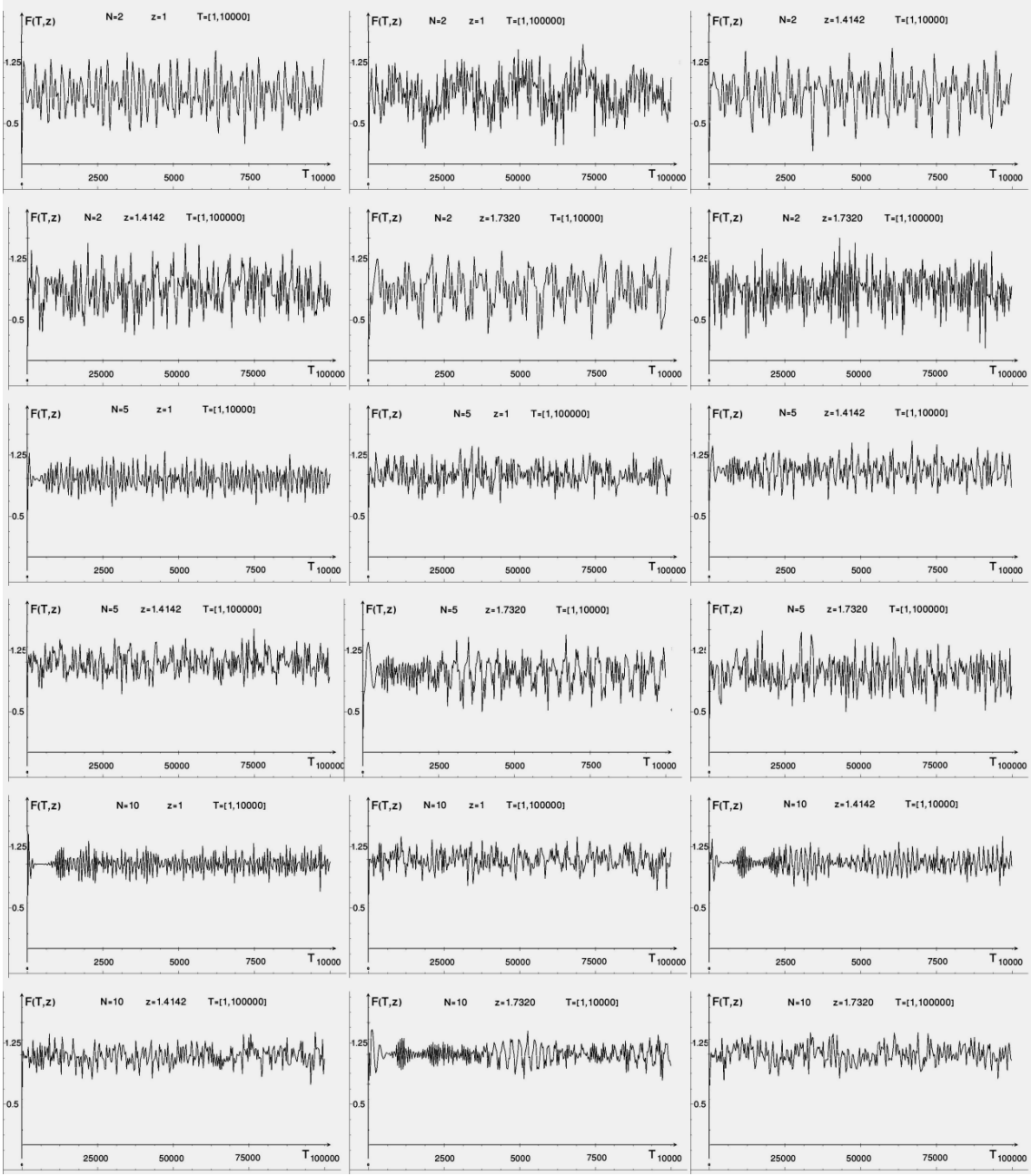

Fig. 4. 2D diagrams of the fluctuation function for different values of N and z in different scaled time intervals. MT are initially in the ground state.

Now, we investigate the behavior of the fluctuation function with the assumption that the MT is first in the excited state. Here $c_e(0) = 1$ and $c_g(0) = 0$. With the substitution of these coefficients we have

$$|\psi_e(t)\rangle = \sum_{n=0}^{\infty} c_n \cos(\lambda t \sqrt{n+1})|n\rangle \qquad (15)$$

$$|\psi_g(t)\rangle = -i\sum_{n=0}^{\infty} c_n \sin(\lambda t \sqrt{n+1})|n+1\rangle \qquad (16)$$

Continuing our calculations for the fluctuation function we find that

$$F(T,z) = \sum_{n=0}^{\infty} e^{-N} \frac{N^n}{n!} \left[ 1 - \sin^2(\sqrt{n+1}T) + \cos^2(e^{-|z|^2/2}\sqrt{n+1}T) \right]$$

We have plotted 3D and 2D diagrams of the fluctuation function in Figures 5 and 6, respectively. As shown before, the maximum of fluctuations is found to be around z=1. Here, we see again these fluctuations around z=1.

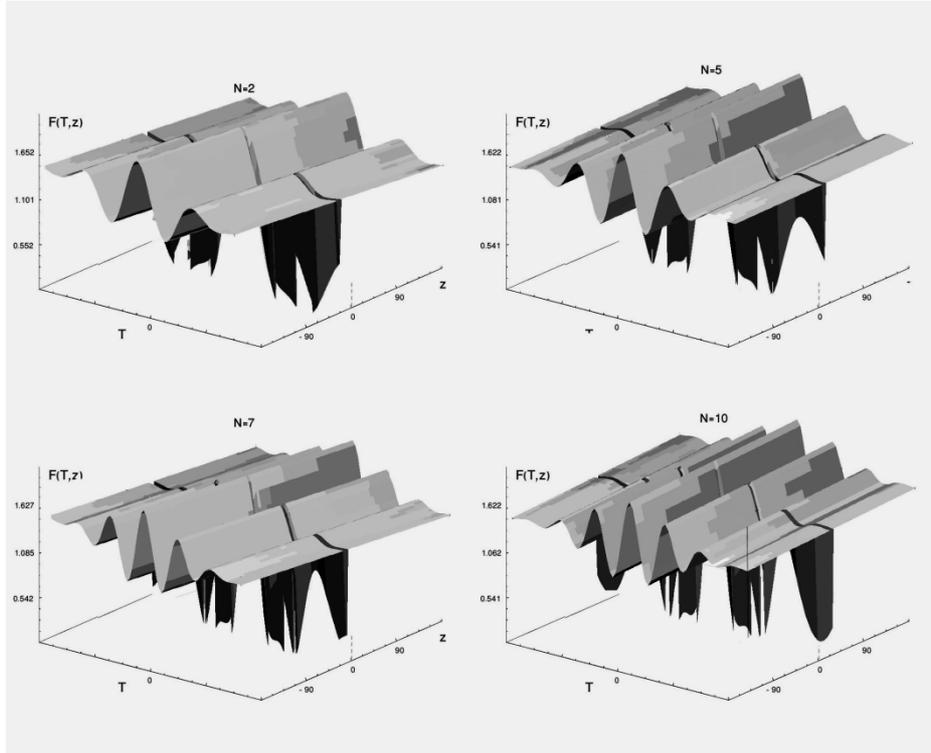

Fig. 5. 3D diagrams of the fluctuation function for different values of biophotons N. In this interaction, MT is initially in the coherent state

Next, we plot F(T,z) again for the above case at different values of its variables.

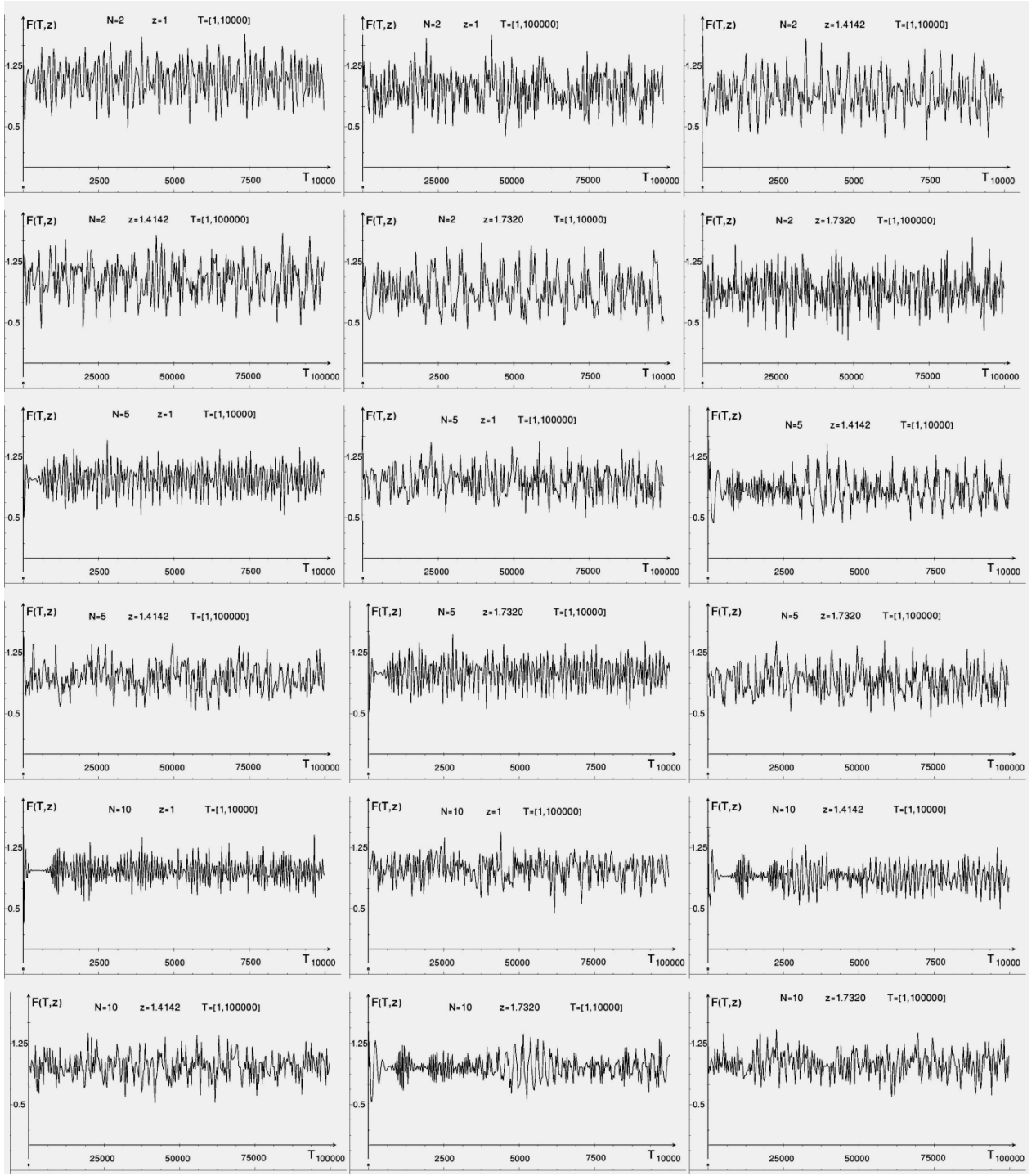

Fig. 6. 2D diagrams of the fluctuation function for different values of N and z. MT is initially in the coherent state.

In Figures 4 and 6 it is seen that the amplitudes of the fluctuation function are decreasing by increasing the number of biophotons which interact with an MT, and vice versa.

**4. Electrical properties of MTs and their effects on the membrane activity**

In previous sections we explained how biophotons can help MTs support coherent states. We have shown the existence of fluctuations between coherent states and normal states for MTs during the interaction with biophotons. Now the question arises how biophotons can affect the electrical activity of membrane via MTs. To answer this question we should investigate how MT activity may affect the electrical activity of the membrane. It is well known that MTs play key roles in the trafficking of neurotransmitters to the synapse. According to Alvarez and Ramirez [5] action potential leads to a decrease of the disassembly rate of MTs. Very recently, Gardiner et al. [25] reviewed the evidence for neurotransmitters regulation (i.e. serotonin, melatonin, dopamine, glutamate, glycine, and acetylcholine) of the MT cytoskeleton. They postulated that MTs may play a direct role in propagating action potentials via conductance of electric current.

Both experimental and theoretical approaches have been used to study electrical signaling along MTs. For experimental investigations, the dual patch-clamp set up was used. In such experiments, electrical data were gathered and taxol-stabilized MTs were shown to behave as biomolecular transistors responding to brief pulses of electric current whose voltage amplitude was in the range of ±200mV [68]. Individual MTs were shown to amplify applied electrical current two-fold indicating a capability for ionic signal propagation that appeared to involve the condensed positive counterion clouds distributed along the length of the MT (where approximately 20 unit charges are present per tubulin monomer). This ionic cloud was attracted to the negative surface charge of the MT retaining its longitudinal mobility. Further measurements and theoretical modeling showed that MTs support nonlinear wave propagation [68]. It was also found that MTs exhibited transistor-like properties. It is worth noting that MTs were also shown to be conductive using an independent experiment involving an electroorientation approach [58]. Intact MTs were demonstrated to have conductance of $157 \pm 7$ mS/m, while MTs treated with subtilisin (which cleaves tubulin's C-termini) lowered it to $96 \pm 6$ mS/m. An argument was put forth that counterions on the surface (many of which interact with the negatively charged C-termini) are responsible for the observed conductance. Ionic conduction along MTs was modeled in terms of a nonlinear electrical circuit using cable equations [69]. Based on these results which indicate that MTs transmit electric signals between distant points within a neuron, the sources of potential information need to be identified in order to understand how processing sensory-based information occurs into a higher cognitive state. The most significant source of input is from the neuronal membrane which contains postsynaptic densities making contacts with other neurons.

In order for actin filaments (AFs) to conduct an electric signal to the MT network inside the neuron, there must be a functional link between these two types of cytoskeleton. There are at least three potential mechanisms through which AFs and MTs could interact: (1) direct physical contact, (2) via various types of linking proteins, and (3) indirectly through signal transduction. AFs are often found in cells forming direct association with MTs. MTs, in turn, frequently migrate correlated with actin bundles [76] as was demonstrated by bual-wavelength fluorescent speckle microscopy analyses. AFs interact with MTs as part of neuronal migration, growth cone development, and neuronal receptor and ion channel transport [38]. While detailed mechanisms governing MT/AF interactions remain an open issue, numerous recently acquired insights reveal a growing number of cross-linker proteins playing important roles.

**Table 1. Major linker proteins of actin filaments with microtubules**

| Cross-linker protein | References |
|---|---|
| CLIP-115 | [38] |
| CLIP-170 | [38],[101] |
| CLASP1 | [38],[93] |
| CLASP2 | [38],[93] |
| Lis1 | [38],[46] |
| EB family | [38],[53] |
| MAP2c | [51],[78] |
| Tau | [78] |

Table 1 summarizes the known cross-linker proteins that bind MTs to AFs. Note that many of these proteins are MT plus-end tracking (so called +TIPs) [38]. Cytoplasmic linker proteins, CLIP-115 and CLIP-170, bind plus-ends of MTs and link them to cargo or to AFs though scaffolding protein intermediaries [101]. CLIP-associated proteins, CLASP1 and CLASP2, represent two additional examples of +TIPs. CLASP2α contains a binding site for actin [93]. Lis1 tethers AF to MTs through interactions with scaffolding proteins [46]. Similarly, the EB family of +TIPs, which include EB1 and EB2, interact with other proteins (including CLASP1/2, CLIP-170) in order to bind MTs to AFs [53]. There are also the microtubule-associated proteins (MAPs): MAP2 and tau, which are known to bind AFs at relatively low affinities [51,78]. Signal transduction molecules, such as calmodulin, and $Ca^{2+}$ and phosphorylation have been found to modulate the ability of MAP2 and tau to bind AFs and MTs. Consequently, in spite of numerous gaps in our understanding of these interactions, it is clear that electric signals arising from synaptic input can reach the internal cytoskeleton using various molecular pathways.

*4.1 An Overview*

We strongly believe that electrodynamic interactions between various cytoskeletal structures, with MTs playing a central role, and ion channels crucially regulate the neural information-processing mechanism. These interactions involve long-range ionic wave propagation along microtubule networks (MTNs) and AFs and exhibit subcellular control of ionic channel activity. Hence, they have an impact on the computational capabilities of the entire neural function. Cytoskeletal biopolymers, most importantly AFs and MTs, constitute the basis for wave propagation, and interact with membrane components leading to a modulation of synaptic connections and membrane ion channels. Association of MTs with AFs in neuronal filopodia guides MT growth and affects neurite initiation [18]. This is seen in neurons by the presence of proteins that interact with both MTs and AFs, as well as proteins that mediate interactions between both types of filaments. For example, MAP1B and MAP2 interact with actin in vitro [64,90] but cross-linking, MAP2 and/or MAP1B is associated with both types of filaments contributing to the guidance of MTs along AF bundles. Direct interaction between AFs and ion channels has been seen and a regulatory functional role has been associated with actin. Thus, it is clear that the cytoskeleton directly and indirectly affects membrane components, in particular ion channels and synapses.

MTs and AFs also interact during the migration of developing growth cones, a process that involves +TIPS, which tend to aggregate at the end of the axon shaft in the area where AFs are highly localized. Additionally, the neurotrophin NGF and the signal transduction molecule GSK-3β act to assure a balance between stable and unstable MTs [28]. The dynamic relationships between AFs and MTs during growth cone migration could be partially controlled by electric signals transmitted between these networks, which is supported by experimental evidence. Applied electric fields were found to guide growth cones of *Xenopus* spinal neurons towards the positive current source and this effect depended on intact AFs and MTs [71].

While learning-related changes in MTs, AFs, MAPs and signal transduction molecules are well documented [103], much less is known about how electric signaling in the MT and AF networks might be involved memory formation processes. Electric signaling by AFs and MTs may play active roles in coincidence detection and storage of spatiotemporal patterns of inputs, and signaling within the cytoskeleton may be particularly critical to information storage over longer time scales than LTP times. The initial route to the MT network could be through the AFs concentrated in the spines. Inputs to arbitrary sites in the neuron can be transmitted from the neuronal membrane to AFs in spines via scaffolding proteins and signal transduction molecules. Electric signals can then be transmitted utilizing AF cross-linker proteins to MTs, and subsequently through MAPs and signal transduction molecules to other MTs in the network. A spatiotemporal pattern inherent to a complex stimulus (or cogneme) can therefore be readily envisaged and indeed mathematically modeled as a convergence of electrical signals to a particular stretch of an MT inside the dendrite.

## 5. Concluding Remarks

There is no doubt that EEG waves are deeply involved with the basic functioning of the brain but the origin and the exact function of EEG has remained a mystery. The EEG waves associated with two distant neurons are strongly correlated and this supports the view that EEG waves are related to the properties of the brain as a coherent quantum system. It is not possible for a scalp EEG to determine the activity within a single dendrite or neuron. Rather, a surface EEG reading is the summation of the synchronous activity of thousands of neurons that have similar spatial orientation, radial to the scalp[2].

Synaptic transmission and axonal transfer of nerve impulses are too slow to organize coordinated activity in large areas of the central nervous system. Numerous observations confirm this view [73]. The duration of a synaptic transmission is at least 0.5 ms, thus the transmission across thousands of synapses takes about hundreds or even thousands of milliseconds. The transmission speed of action potentials varies between 0.5 m/s and 120 m/s along an axon. More than 50% of the nerves fibers in the corpus callosum are without myelin, thus their speed is reduced to 0.5 m/s. How can these low velocities (i.e. classical signals) explain the fast processing in the nervous system? We believe that quantum theory is able to explain some of the above mysteries. As an example, recently it has been shown theoretically that the biological brain has the possibility to achieve large quantum bit computing at room temperature, superior when compared with the conventional processors [60].

---

[2] Updated from : http://www.reference.com/browse/electro-encephalogram, 13 November 2009.

Neuroscientists or brain specialists record the EEG diagrams of their patients when their eyes are closed, because when their eyes are open the amplitudes of diagrams dramatically would be reduced and they cannot diagnose changes in the amplitudes. In Figure 7 we represent the EEG diagrams of a person with different situations as explained in the caption to the figure. When the eyes are opened, the number of interacting photons with the biomolecules becomes high and the amplitudes of EEG become low. It is clear when the intensity of incident light is high; it must produce more action potentials than low intensity light. In classical physics, it is argued that this situation is because the superposition of several waves tends to decrease the amplitude of the total wave. This argument ignores the intensity of the incident light entering the eye. This classical argument also cannot explain when and how the synchronization happens in order to decrease the amplitude of EEG.

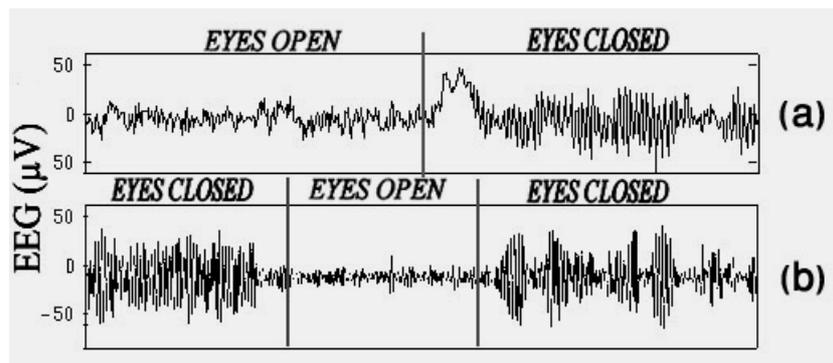

Fig. 7. A schematic alpha-EEG diagram of a person when (a) the eyes are opened and then he/she closes the eyes (b) the eyes are closed and then she/he opens his eyes and then closes them again.

Our argument is based on the interaction of (bio)photons and MTs inside the neurons. When the eyes start to be opened, the incident numbers of photons into the eyes increases and according to the diagrams in Figures 4 and 6 the amplitudes are reduced. Also, the amount of incident photons can increase the amount of biophoton production inside the neurons, since the visible light can be produced inside the brain in the form of biophotons [37,49,50,100] and this interaction is not only limited to the external incident photons with MTs. So far we have not been able to find an exact relation between the EEG diagrams and the fluctuation function, but the synchronous and coherent vibrations of billions electric dipoles of biomolecules cannot be ignored in the EEG diagrams. MTs are particularly numerous in the brain where they form highly ordered bundles and are the best candidate for long coherence and large synchrony [32]. The argument for connection between Alpha-EEG diagrams and MTs activity is their similar behavior in increasing and decreasing of amplitudes of fluctuation function for MTs and potential difference in EEG in response to the intensity of photons. This similarity during opening and shutting of the eyes indicates a significant relation between the EEG diagrams and the fluctuation function.

**Acknowledgments**

Majid Rahnama thanks the Kerman Neuroscience Research Center (KNRC) for supporting this work, and also István Bókkon gratefully acknowledges the support he has received from the BioLabor (www.biolabor.org), (Hungary) during this research. His URL: http://bokkon-brain-imagery.5mp.eu. Jack Tuszynski acknowledges support from NSERC (Canada) for his research. Michal Cifra acknowledges financial support from Czech Science Foundation, P102/10/P454.**References**

[1] Adamo AM, Llesuy SF, Pasquini JM, Boveris A, Brain chemiluminescence and oxidative stress in hyperthyroid rats, *Biochem J* **263**:273–277, 1989.
[2] Albrecht-Buehler G, Autofluorescence of live purple bacteria in the near infrared, *Exp Cell Res* **236**:43–50, 1997.
[3] Albrecht-Buehler G, Cellular infrared detector appears to be contained in the centrosome, *Cell Motil Cytoskeleton* **27**:262–271, 1994.
[4] Albrecht-Buehler G, Changes of cell behavior by near-infrared signals, *Cell Motil Cytoskeleton* **32**:299–304, 1995.
[5] Alvarez J, Ramirez BU, Axonal microtubules: their regulation by the electrical activity of the nerve, *Neurosci Lett* **15**:19–22, 1979.
[6] Bajpai RP, Coherent nature of biophotons: Experimental evidence and phenomenological model, pp. 323–339, in Chang J, Fisch J, Popp FA (eds.), *Biophotons*, Kluwer Academic Publishers, Dordrecht, 1998.
[7] Bajpai RP, Coherent Nature of the Radiation Emitted in Delayed Luminescence of Leaves, *J Theor Biol* **198**:287–299, 1999.
[8] Batyanov AP, Distant optical interaction of mitochondria. *Bull Exp Biol Med* **97**:740–742, 1984.
[9] Bischof M, Biophotons - The light in our cells, *J Optom Photother*, March, 1–5, 2005.
[10] Bókkon I, Phosphene phenomenon: A new concept, *BioSystems* **92**:168–174, 2008.
[11] Bókkon I, Salari V, Tuszynski J, Antal I, Estimation of the number of biophotons involved in the visual perception of a single object image: Biophoton intensity can be considerably higher inside cells than outside. *J Photochem Photobiol B* **100**:160–166, 2010.
[12] Bókkon I, Visual perception and imagery: a new hypothesis, *BioSystems* **96**:178–184, 2009.
[13] Chiarugi P, Reactive oxygen species as mediators of cell adhesion, *Ital J Biochem* **52**: 28–32, 2003.
[14] Chiarugi P, Src redox regulation: there is more than meets the eye, *Mol Cells* **26**:329–337, 2008.
[15] Chwirot BW, Ultraweak luminescence studies of microsporogenesis in Larch, pp. 259–285, in Popp FA, Li KH, Gu Q (eds.), *Recent advances in biophoton research and its applications*, World Scientific, Singapore, River Edge, NJ, 1992.
[16] Cifra M, Pokorný J, Havelka D, Kučera O, Electric field generated by axial longitudinal vibration modes of microtubule, *BioSystems* **100**:122 – 131, 2010.
[17] Cilento G, Photobiochemistry without light, *Experientia* **44**:572–576, 1988.
[18] Dehmelt L, Halpain S, Actin and microtubules in neurite initiation: Are MAPs the missing link? *J Neurobiol* **58**:18–33, 2004.
[19] Deriu MA, Soncini M, Orsi M, Patel M, Essex JW, Montevecchi FM, Redaelli A, Anisotropic Elastic Network Modeling of Entire Microtubules, *Biophys J* **99**:2190 – 2199, 2010.
[20] Dröge W, Free Radicals in the Physiological Control of Cell Function, *Physiol Rev* **82**:47–95, 2002.
[21] Faber J, Portugal R, Rosa LP, Information processing in brain microtubules, *BioSystems* **83**:1–9, 2006.
[22] Fröhlich H, Long range coherence and energy storage in biological systems, *Int J Quantum Chem* **2**:641–649, 1968.
[23] Fröhlich H, Long range coherence and the action of enzymes, *Nature* **228**:1093, 1970.
[24] Fröhlich H, The extraordinary dielectric properties of biological materials and the action of enzymes, *PNAS* **72**:4211–4215, 1975.
[25] Gardiner J, Overall R, Marc J, The microtubule cytoskeleton acts as a key downstream effector of neurotransmitter signaling, *Synapse* **65**:249–256, 2011.
18